# Matching the Unmatchable.
# Complexity Theory and Quantum Theory

# Joigner ce qui est unjoignable.
# Théorie de la Complexité et théorie quantique

Carlos Eduardo Maldonado[*]

*Epistemology of Complexity*
*Épistémologie de la Complexité*

**Abstract**
This is a philosophical paper. It claims that there is a gap to be filled in the relationship between complexity theory (CT) and quantum theory (QT). This gap concerns two very distinct understandings of time. The paper provides the ground for filling up such gap. Indeed, most works on complexity at large focus on the macroscopic world, leaving behind the importance of the microscopic world. This paper specifies what both worlds consist of, and argues that a solid account of the world, i.e. complexity, must necessarily take into account both dimensions of reality.

**Resumé**
Celui-ci c'est un article philosophique. Il soutient qu'il y a un gap qui doit être rempli dans la relation entre la théorie de la complexité (TC) et la théorie quantique (TQ). Ce gap consiste en ceci qu'on trouve deux compréhensions parfaitement différentes vis-à-vis le temps. Cet article fourni les bases pour remplir le gap. Il est certain, la plupart des travaux concernant la complexité se concentrent sur le monde macroscopique, et laissent de côté l'importance du monde microscopique. Ce texte rend explicite en quoi consiste chacun des deux mondes et affirme qu'un compte-rendu solide du monde, voire de la complexité doit nécessairement tenir compte des deux dimensions de la réalité.

**Keywords**
Complexity theory; science; time; quantum physics

---
[*] Full Professor, Universidad del Rosario (Bogotá, Colombia). Ph.D. in Philosophy from KULeuve (belgium). Visiting Positions: as Visiting Scholar (U. of Pittsburgh), Visiting Reserac Professor (The Catholic University of America, Whashington, D.C.), Academic Visitor-Visisiting Scholar (University of Cambridge, UK). He holds an honorary doctorate from the University of Timisoara (Rumania). He works on complexity science at large. Email: carlos.maldonado@urosario.edu.co. School of Political Science and International Relations, Universidad del Rosario, Calle 12 No. 6-25, Bogotá, Colombia.

# 1. Introduction

Whereas the followers of complex thinking (*la pensée complexe*) have barely mentioned one word on quantum physics, not to mention quantum theory (QT), complexologists who work on complexity science have undergone some works on the issue. The truth is that there is an increasing complexity also in the quantum world (Anders and Wiesner, 2011). Being as it might be, the truth is that nearly all works on complexity at large focus on the macroscopic world, leaving the microscopic universe far behind. This is a serious problem.

Quantum theory is by and large the most verified, confirmed, falsified, robust theory in the world. It has been tested to the eleventh decimal, and one third of the world economy entirely depends on quantum science. No explanation of the world and the universe is tenable that does not cross, encompass, or takes into account somehow the principles of (QT).

I here talk about quantum theory for, as it is well known, there is not just quantum physics, but also quantum chemistry, quantum biology, and to a large extend, technologies based on quantum principles or behaviors (Hida *et al.*, 2004). Hence, the appropriate concept becomes quantum theory. There, though, is the wrong belief that quantum phenomena are microscopic. This forgets that there are, as it happens, also macroscopic quantum behaviors, such as the condensates Einstein-Bose, all phenomena of superconductivity, and superfluidity, or also all the studies and experiments on teletransportation, for instance. Thus, quantum studies are not reduced to the atomic scale – not any longer. Because there is not enough work on (QT) on the behalf on many complexologists, it is enough to say that a good general view of the quantum world encompasses three aspects, thus: quantum mechanics, wave mechanics, and entanglement.

There are some works on the relationship between complexity theory and quantum theory. However, most of them deal with just one single branch of complexity, namely: computational complexity (Cai *et al*, 2015). Important and suggestive as it is, that remains as yet as only one of the manifold faces of research on complexity. What is sure is that it is, *faute de mieux*, a novel and unexplored thrilling subject.

Here I shall argue that (QT) deals with microscopic time scales, and that these scales are increasingly the real time – in almost all domains of our current world. In contrast, as they are usually studied, complex phenomena fall within the range of macroscopic timescales. To be sure, the aim of this text consists in matching the two dimensions of reality and the universe namely the macroscopic and the microscopic time measures of the world. To say the least, there is a number of shared aspects common to both complexity and quantum science (Larson, 2016). Table No. 1 shows the two time scales of the universe.

Table No. 1: Macroscopic and microscopic time scales

| Macroscopic time scales | Microscopic time scales |
|---|---|
| Second = 1/60 minute | Mili = $10^{-3}$ |
| Minute = 1/60 hour | Micro = $10^{-6}$ |
| Hour = 60 minutes | Nano = $10^{-9}$ |
| Day = 24 hours | Pico = $10^{-12}$ |
| Month = 30/31 days | Femto = $10^{-15}$ |
| Year = 365 days | Atto = $10^{-18}$ |
| Million years = $10^6$ years | Zepto = $10^{-21}$ |
| Billion years = $10^9$ years | Yocto = $10^{-24}$ |

Source: the author

When compared to the microscopic time scales, macroscopic time scales are really slow. Microscopic time scales are increasingly fast. This topic has been worked out in (Maldonado, Gómez-Cruz (2014).

**2. Matching the Unmatchable**

Plainly said in just one word, complexity is about time. Time explains, serves for the rationale, and grounds complexity at large, no matter what understanding one may have about complex systems. In other words, complexity is about the importance and meaning of the arrow of time. This means, that, as it is well known, the past is qualitatively different from the future. I. Prigogine sets out the first building stone for the sciences of complexity when, as the Sweden Academy of Sciences stated when he was awarded the Nobel Prize in 1997, because he introduced in science what sciences until then did not have: time. It is the arrow of time what properly allows us to speak not just about complexity but even better, about increasing complexity. A right understanding of complex systems means reckoning that they are systems of *increasing* complexity (not just, complex, period).

Now, from this standpoint, the trouble with quantum theory is that it does not know of time at all, certainly not in the sense complexologists talk about. The best one can say about time in the framework of (QT) is that time is present. More accurately, time is expressed in the principle of superposition: time is (pure) present and in present all the possibilities exist at the same time. Roughly said, Schrödinger's cat is both dead and alive, at the same time – well before one looks into the box. Beyond the principle of superposition, nearly all equations in quantum physics are linear, which means, they either do not know about the arrow of time, or else, they do not care about time as an arrow. Quantum mechanics does not know about past and future, and their differences.

Can we match both comprehensions – namely the arrow of time of complexity, and the absence of an arrow of time where time is pure present, superpositions – in (QT)? This is the crux of this paper. I shall claim that both stances can and should be matched.

Beyond the particular discussions on the philosophy of science, for instance concerning the debate between internalism and externalism, complexity theory (CT) can be rightly understood beyond the Copenhagen debate – i.e. beyond the debate about the importance of the observer. It should be noted that a classical debate regarding the nature of complexity falls within the frame of the epistemological Copenhagen debate. As a matter of fact, the interpretation of Copenhagen is but just one of more than fourteen interpretations about quantum mechanics.

Nature, society, and the universe as a whole are increasingly becoming complex. A good amount of the literature has already highlighted this fact (Chaisson, 2001; Linewaever *et al*., 2013). The touchstone here concerns the relationship between increasingly complexity and entropy (Tsallis, 2016). Thereafter, the arrow of time sheds constantly new lights on the structure and the dynamics of the world. Complexity has been adequately been said as a physics of becoming (Prigogine) and as creative becoming universe (Kauffman).

It is my contention that superposition: a) triggers the arrow of time, or also b) is at the basis of the arrow of time that triggers the evolution of the universe and life. Indeed, in many senses, it has been repeatedly said that quantum states lie at the bottom, or on the ground, of all explanations of reality. In other words, no good account of society, the world and life is feasible that does not necessarily take into account (QT).

The trouble, though, as mentioned is that in either of its expressions (QT) does not know about the arrow of time. The core of the question here concerns coherence time, namely how long can a quantum state live. A right understanding of the issue is expressed in both the (principle of quantum) coherence and re-coherence. Quantum states live shortly, for there arises the principle or problem of decoherence. That said, it should be noted that there is no absolute (or 100%) quantum state as well as there is neither a pure (or 100%) classical state. Most of the universe occurs in or as quantum-like states. I leave here aside for reasons of space the discussion about closed and open quantum states.

There are no complex systems *in* time (*à la* Newton or Kant). On the contrary, time itself complexifies systems, behaviors, or phenomena, but it does not equally or evenly complexify everything at the same time. The evolution of time knows of a variety of speeds, landscapes, and attractors.

The arrow of time happens as bifurcations, or also as a continuous opening up or gaining degrees of freedom. In other words, evolution does not concern just the passing from the prior to the posterior, the inferior to the higher in any sense that it may be taken. Rather, evolution consists of punctuated equilibrium – inflections. The translation of those concepts in physics is: first order and second order transitions.

The coherence-decoherence state is the moment when a superposition is broken down, so to speak, and the arrow of time is triggered. Hence, begins the big picture of complexification of reality. The marvelous point, though, is that the coherence-decoherence dynamics takes place continuously, unceasingly. The story of the Big-Bang did not happen just once, but it is continuously taking place on and on. We do live in a creative universe, indeed – and life, the most appealing phenomena in our universe, is a never-ending-and-continuously-emerging process. Time is evolution.

Microscopic times are greatly faster when contrasted with macroscopic time scales. Moreover, as it happens along the history of science and technology, we can safely say that *real* time has come to be discovered as the microscopic times. Some examples of this are: attention, health-disease, financial time, security time, the time of the processes beneath the cell such as protein folding – and several others (Melkikh, 2013; Maldonado, Gómez-Cruz, 2014). The entire history of spearhead science and technology – whence, also the history of culture as we have come to experience it – lies on these acknowledgements.

The relationship between both scales is such that we have come to "see" the microscopic time scales on the macroscopic processes, structures and dynamics – but sometimes it might be already too late. Without being reductionist, a safe scientific and philosophical account of time must reckon that time arises, emerges, or is nurtured from the microscopic scales, on to the macroscopic levels and layers.

The world is nurtured, so to speak, from the quantum dimension. The scientific and philosophical trouble is not on this side of the equation, but on how and why the classical world originates. Nonetheless, the idea is not to be accepted that quantum states set out the materials or "stuff" for our universe, for there is also a continuous recoherence as it has been experimentally observed (Bouchard *et al*., 2015).

I would like to go on as to say that the two big time dimensions of the world do not just co-exist, but, moreover, they are entangled. The entanglement can be so grasped that what emerges in the microscopic world is viewed in the macroscopic world – as a macroscopic phenomenon and its dynamics, precisely. But what takes place in the macroscopic world is observed elsewhere in the microscopic level. As in many other observations on both worlds, the main limitation to-date concerns the technology – and the mindset we have that do not allow us to fully grasp the interweavement; even better, the entanglement.

Originally and fundamentally reality is about a large set of possibilities that, from the outset, have equal possibilities of being accomplished. It is so not that any possibility can be fulfilled, but rather that all possibilities exist at the same time. The photon does run along all possible paths before reaching a screen. The two-slit experiment by Thompson remains a holy grail for understanding quantum phenomena (Woesler, 2007).

When quantum coherence is broken down time appears, as we know it, namely bearing an arrow that qualitatively differentiates past from future. As it has been sufficiently been brought out by the sciences of complexity, however, time does not happen linearly. Linearity is a particular case of larger scope, nonlinearity.

Finally, last but not least, it should be always reminded that within, and thanks to, the framework of (QT), the observation, i.e. the very act of measurement does both create and perturb the observed object. In other words, there is neither a subject nor an object – and their relationships, but an entangled relationship. My claim here is that such entanglement is exactly the point where both the microscopic and the macroscopic timescales are linked and differentiated at the same time. We can intuitively say that very much as the observation, i.e. the measurement both creates and disturbs the observed object, so too the two dimensions of time are both united and differentiated in their entanglement. The differentiation occurs with quantum decoherence.

## 3. Conclusions

A right understanding of our world and the universe is not just about phenomena, processes, structures, and dynamics as such. It is, rather, about the weave of different time scales. Roughly said, we can assess that the two basic layers of time are macroscopic and microscopic. A good account of the complexity of the world pertains the interweavement of macroscopic *and* microscopic behaviors and systems. Diagram No. 1 illustrates the basic idea of this paper.

**Diagram No. 1: Timescales and the relationship between QT and CT**

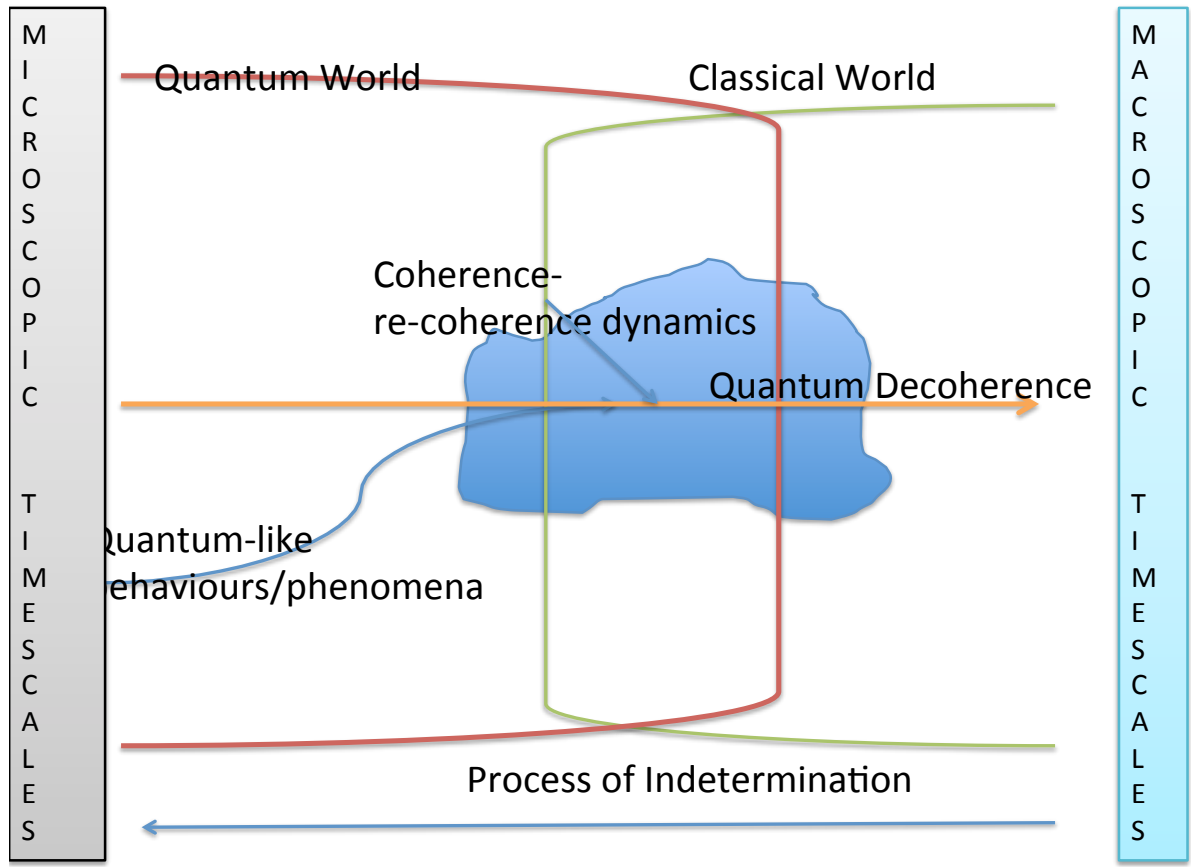

Source: the author

Generally said, complex systems are, all in all, open systems. Now, the *conditio sine qua non* for seeing and coping with open systems consists in being open-minded; some have even come to soundly speak about quantum cognition (Conte, Norman, 2016). Here I have argued that relating the until-now un-relatable (CT) and (QT) does firmly contribute to see and live in a complex world. Complexity is not without quantum behaviors.

Most workings relating on the relationship between (QT) and (CT) focus on information and computation (Campbell-Borges and Castilho Piqueira, 2012). There are, to be sure, good arguments for so doing. And yet, the informational and computational concerns do not entirely encompass the trouble about complexity, even if they shed good lights.

There is a call for a quantum complexity theory (Bernstein and Vazirani, 1997; Gruska, 2005; Baskaran, 2011). The ground is served for seriously talking about quantum complex systems (Burghardt and Buchleiter, 2015). Such a call, however, remains to-date the call of a lone wolf, so to speak – for not many scholars and researchers have seriously dig into that

expression – as suggestive as it remains. The general frame for the call has been and remains informational and computational. Sound doubts can be raised about whether the computational scope is sufficient for dealing with complexity as such – namely complexity at large. This is the reason why the scope of this paper has been philosophical, meaning calling for a larger picture – the *big picture* of life, our world, and nature.

Complexologists at large still remain victims of the macroscopic worldview. Such an attitude is highly limited and even dangerous. A much richer and wider scope is possible thanks to the quantum states. As it follows from the above, the gap between (CT) and (QT) is filled up.